# Ion-Size Effect at the Surface of a Silica Hydrosol[1]

Aleksey M. Tikhonov*

*P. L. Kapitza Institute for Physical Problems, RAS 119334, Kosygina 2, Moscow, Russia*

*E-mail: tikhonov@kapitza.ras.ru

**August 28, 2008**

ABSTRACT

The author used synchrotron x-ray reflectivity to study the ion-size effect for alkali ions ($Na^+$, $K^+$, $Rb^+$, and $Cs^+$), with densities as high as $4 \times 10^{18} - 7 \times 10^{18}$ m$^{-2}$, suspended above the surface of a colloidal solution of silica nanoparticles in the field generated by the surface electric-double layer. According to the data, large alkali ions preferentially accumulate at the sol's surface replacing smaller ions, a finding that qualitatively agrees with the dependence of the Kharkats-Ulstrup single-ion electrostatic free energy on the ion's radius.

The participation of small inorganic ions in a variety of surface phenomena at air-liquid and liquid-liquid interfaces has practical significance for many applications.[1-3] Many previous authors discussed the ionic distributions, surface tension, image forces, and single-ion free energy at the surface of an electrolyte solution.[4-15] Also, for decades molecular-dynamics simulations were extensively used to explore molecular structure and ion-specific effects at liquid surfaces.[16-18] However, here synchrotron x-ray scattering is proving invaluable as it offers information on a liquid's surface structure at a microscopic level, giving details that cannot be acquired by measuring macroscopic characteristics, such as surface tension, interfacial capacitance, or surface potential.[19-29] In this paper, I discuss my findings using synchrotron x-ray reflectivity to elucidate the ion-size effect for alkali ions ($Na^+$, $K^+$, $Rb^+$, and $Cs^+$) elevated above the surface of a colloidal solution of silica nanoparticles by the field of the surface electric-double layer.

In the traditional Wagner-Onsager-Samaras approximation, ions are treated as point charges.[8, 9] The major difficulty of this approach concerns the divergence of the free energy of a point charge at a flat interface between two dielectric media. Kharkats and Ulstrup resolved this problem by assuming that the ion has a finite size.[10] According to them, in continuous media approximation, the following is the free energy, $F(z)$, of a spherical charge, $q$, with radius, $a$, at the boundary between two dielectric media imbedded within a spherical cavity:

$$F(0 \leq z \leq a) = \frac{q^2}{32\pi\varepsilon_0\varepsilon_2 a}\left[2 + \frac{2z}{a} + \left(\frac{\varepsilon_2-\varepsilon_1}{\varepsilon_1+\varepsilon_2}\right)\left(4 - \frac{2z}{a}\right) + \left(\frac{\varepsilon_2-\varepsilon_1}{\varepsilon_1+\varepsilon_2}\right)^2\left(\frac{(1-z/a)(1-2z/a)}{1+2z/a} + \frac{a}{2z}\ln\left\{1+\frac{2z}{a}\right\}\right)\right] + \frac{q^2}{16\pi\varepsilon_0\varepsilon_1 a}\left(\frac{2\varepsilon_1}{\varepsilon_1+\varepsilon_2}\right)^2\left(1-\frac{z}{a}\right); \quad (1)$$

and,

$$F(z \geq a) = \frac{q^2}{32\pi\varepsilon_0\varepsilon_2 a}\left[4 + \left(\frac{\varepsilon_2-\varepsilon_1}{\varepsilon_1+\varepsilon_2}\right)\frac{2a}{z} + \left(\frac{\varepsilon_2-\varepsilon_1}{\varepsilon_1+\varepsilon_2}\right)^2\left(\frac{2}{1-(2z/a)^2} + \frac{a}{2z}\ln\left\{\frac{2z+a}{2z-a}\right\}\right)\right], \quad (2)$$

where $\varepsilon_0$ is the dielectric permittivity of the vacuum, $\varepsilon_1$ and $\varepsilon_2$ are the dielectric permittivities of the bottom- (water $\varepsilon_1 \approx 80$) and top- (air $\varepsilon_2 \approx 1$) phases respectively, and the *z*-axis is directed normal to the surface (oriented by gravity) towards the top phase. We obtain the electrostatic free energy of the ion in the water ($z \leq 0$) from Eqs. (1) and (2) by exchanging $\varepsilon_1 \Leftrightarrow \varepsilon_2$. Several authors revised and discussed Eqs. (1) and (2)[12, 14, 16] For example, Markin and Volkov used them to explain the dependence of the surface tension of aqueous electrolyte solutions on ionic radii.[14]

In general, an alkali ion in aqueous media is repelled from the water's surface (towards water's bulk) by its electrical image. The total thickness of the transition layer at the surface of concentrated aqueous solution of simple inorganic ionic salt (for example, CsCl) is less than 1 nm wide.[28] However, the larger the ion's radius the weaker it interacts with the boundary, although this is important only in the very narrow interfacial region, ~ *2a,* about as wide as the size of the ion, above the water's surface (Fig. 1a). At a distance of several ion radii from the surface, the ion interacts with the boundary as a point charge: For the Na$^+$ radius $a \approx 1.2$ Å; for K$^+$ $a \approx 1.5$ Å; for Rb$^+$ $a \approx 1.7$ Å; and, for Cs$^+$ $a \approx 1.8$ Å.[30, 31]

The solid lines in Fig. 1b depict the deviation of the single-ion Kharkats-Ulstrup free energies $F_M(z)$ of these monovalent alkali ions M$^+$ (= K$^+$, Rb$^+$, Cs$^+$) from the energy of Na$^+$, $F_{Na}(z)$, at the air-water interface; the dashed line represents the difference between $F_{Cs}(z)$ and $F_K(z)$. On the one hand, at $z < 0$ $\Delta F <$ 0.03 eV (~$k_B T$, at *T*=298 K and $k_B$ is Boltzmann's constant) is small, featureless, and associated mostly with the difference in the Born solvation energies, $F_M(-\infty)$, of the ions in water. On the other hand, at $z \geq 0$, these curves display minima as deep as 0.05-0.1 eV ($2k_B T - 4k_B T$) at ~ 2 Å above the surface of the water: larger alkali ions preferentially accumulate there, replacing smaller ions (ion-size effect). Usually, this effect is unimportant at room temperature because, for example, the elevation of Na$^+$ ~ 1 Å above the water's surface, is associated with a significant energy barrier ~ 2.5 eV; overcoming it would require very specific boundary conditions, *viz.,* an interfacial electric field with the strength $>10^9$ V/m. A field of such strength, which is impossible to obtain in an electrolytic capacitor, is common at the surface of a silica hydrosol that is strongly polarized by the forces of electrical imaging.[32, 33]

Previous x-ray reflectivity and grazing-incidence diffraction data for NaOH-stabilized and Cs-enriched suspensions with monodispersed 5-, 7- and 22-nm silica particles suggest that a four-layer model can describe the structure of the hydrosol's surface (see Fig. 2).[32-35] The top two layers in Fig. 2 reflect the adsorption of alkali ions, i.e., a low-density layer 1 of suspended (elevated) ions, and a layer 2 of space charge with the surface density of $Na^+$ $\Phi_{Na} \approx 8 \times 10^{18}$ m$^{-2}$. The former is inhabited either by $Na^+$ and/or $Cs^+$ ions, depending on the bulk concentration of cesium, $c_{Cs}^+$, in the hydrosol with roughly one water molecule per ion. On the contrary, the space charge layer formed by the hydrated ions with ten $H_2O$ molecules per alkali ion. The depleted layer 3 with low electrolyte concentration (~ 10-20-nm thick) separates these layers from the anionic colloidal particles: its density roughly equals the density of bulk water at normal conditions, $\rho_w$ (= 0.333 e$^-$/Å$^3$). Finally, the thickness of layer 4 is the same as the diameter of the colloidal nanoparticles in the sol; the concentration of particles in the loose monolayer is up to twice as high as in the bulk.

The pronounced width of the transition region (~ 20 - 50 nm) at the hydrosol's surface reflects the extremely large difference between the forces of electrical imaging for nanoparticles and the monovalent alkali ions. In fact, it is comparable to the Debye screening length in the solution, $\Lambda_D = \sqrt{\varepsilon_0 \varepsilon_1 k_B T / (c^- N_A e^2)} \approx 10-100$ nm, wherein $N_A$ is the Avogadro constant; $e$ is the elementary charge; and, $c^-$ is the bulk OH$^-$ concentration ($c^- \approx 10^{-3} - 10^{-5}$ mol/L at pH = 9 – 11).

With increasing $c_{Cs}^+$, the density of layer 1 rises (Cs$^+$ replaces Na$^+$ in the layer) and then saturates ($c_{Cs}^+$ > 0.1-0.2 mol/L), so that for Cs-enriched sols with 5-, 7-, and 22-nm particles the reported surface density of Cs$^+$ in layer 1 was the same and reached as high as $\Theta_{Cs} \approx 3 \times 10^{18}$ m$^{-2}$. Dissimilarly, the densities of the layers 2, 3 and 4 virtually do not depend on, $c_{Cs}^+$. The strength of the electric field (normal to the surface) of the space charge layer 2, which supports the elevated ions in layer 1, is as high as $\sim \Phi_{Na} e / \varepsilon_0 \varepsilon_1 \sim 10^9\text{-}10^{10}$ V/m; hence, in the Kharkats-Ulstrup theory, the effect of the preferable adsorption of Cs$^+$ in layer 1 can be considered as a manifestation of the ion-size effect for the suspended ions. Furthermore, the energy barrier for cation to cross the hydrosol's surface should be smaller than shown in Fig. 1 due to small dialectic permittivity of water in the surface electric double layer. For example, at the surface charge density $\Phi_{Na} e \sim 1$ C/m$^2$, $\varepsilon_1 \sim 3$ (see, Figure 1 in Ref. 36).

I systematically studied the same effects on the density of layer 1 at the surfaces of monodispersed suspensions of 22-nm silica particles enriched by different alkali ions (K$^+$, Rb$^+$, Cs$^+$). To ensure the saturation of layer 1, I chose a bulk concentration of alkali metals, $c^+$, in the hydrosols that was significantly larger than the sodium concentration ($c^+ >> c_{Na}^+ \approx 0.06$ mol/L). Following the methods in Ref. [34], the solutions were prepared by mixing either mechanically or in an ultrasonic bath (Branson 2510) a 1:1 (by weight) solution of alkali metal hydroxide MOH in deionized water (Barnstead UV) with an NaOH-stabilized sol of 22 nm silica particles (~ 30% of SiO$_2$ by weight).[37] The total alkali concentration in the sol ranged from 0.7 – 0.9 mol/L.[38] The size of silica particles in the hydrosol was selected specifically to facilitate interpretation of x-ray reflectivity data: the larger the particles, the smaller their contribution to reflectivity at high incident angles. This relationship is evidenced both by the wide surface-normal structure of the 22-nm particle's sol and the high surface roughness of the loose monolayer (see Fig. 2). At room temperature, these suspensions (pH < 11.5) remain liquid in a closed container for at least one month.

I carried out all x-ray reflectivity measurements at beamline X19C, National Synchrotron Light Source, Brookhaven National Laboratory, employing a monochromatic focused x-ray beam ($\lambda = 0.825 \pm 0.002$ Å).[39] Liquid samples with a planar surface were studied in a ~ 50 ml capacity Teflon dish with a 100-mm-diameter circular interfacial area placed inside an air-tight single-stage thermostat and mounted above the level of water in a bath (~200 mm diameter); the bath served as a humidifier in the thermostat. Normally, the samples were equilibrated at $T = 298$ K for at least twelve hours. Reflectivity was measured with the detector's vertical angular acceptance at $\Delta\beta = 6.8\times10^{-2}$ deg (twice as high than in Ref. [34]) and its horizontal acceptance at ~ 0.8 deg.

Fig. 3 shows x-ray surface reflectivity, $R(q_z)$, of the sols as a function of wave-vector transfer, $q_z = (4\pi/\lambda)\sin(\alpha)$, where $\alpha$ is an incident angle (see insert in Fig. 3). It is normalized to the Fresnel function, $R_F(q_z)$ that is, the reflectivity from a sharp surface with no structure. The structure factor, $R(q_z)/R_F(q_z)$, consists of two parts: the low $q_z$-part ($q_z < 0.05$ Å$^{-1}$) is associated with a surface-normal distribution of nanoparticles; at $q_z > 0.1$ Å$^{-1}$ the surface-normal structure depends strongly on the alkali-metal composition of the sols. The oscillations of reflectivity at $q_z > 0.1$ Å$^{-1}$ depend on the sample's equilibration history: usually, they were stronger when the hydrosol's temperature was ~ 30 K (pH~13) higher than the room temperature in the beginning of the equilibration (open symbols in Fig. 3). This effect is probably due to the narrowing of the surface-electric double layer at pH~13, so that more alkali ions are available for adsorption (in the equilibrium pH < 11.5). Once the sample was equilibrated in the thermostat, the reflectivity curves were reproducible within error bars for several days.

Both Parratt formalism (see Ref [34] for details) and the first Born approximation were used to obtain information about the distribution of adsorbed ions from x-ray reflectivity values.[40-42] The former also generates data about the surface-normal distribution of nanoparticles from the reflectivity near the angle of total reflection of a hydrosol's surface, $\alpha_c$. However, when multi-photon scattering is unimportant (usually at $\alpha > 3\alpha_c$) the first Born approximation relates reflectivity to the electron-density gradient normal to the interface, $\langle d\rho(z)/dz \rangle$, averaged over the interfacial plane as the following:

$$\frac{R(q_z)}{R_F(q_z)} \approx \left| \frac{1}{\rho_b} \int_{+\infty}^{-\infty} \left\langle \frac{d\rho(z)}{dz} \right\rangle \exp(iq_z z)dz \right|^2 , \qquad (3)$$

where $R_F(q_z) \approx \left(q_z - [q_z^2 - q_c^2]^{1/2}\right)^2 / \left(q_z + [q_z^2 - q_c^2]^{1/2}\right)^2$ is slightly different for each sol since $q_c \approx (4\pi/\lambda)\alpha_c$ is defined by the angle of total reflection $\alpha_c = \lambda\sqrt{r_e\rho_b/\pi} \approx$ 0.09 deg, where $r_e = 2.814 \times 10^{-5}$ Å is the electron's Thomson's scattering length. The bulk electron-densities of the sols, $\rho_b$, are established from their densities and known chemical compositions (Table I).

At $q_z > 0.1$ Å$^{-1}$, only three interfaces (top two layers with adsorbed ions) contribute to reflectivity since $\sigma_3 \sim \sigma_4 > 30$ Å (see Ref. 34). Then, for the slab model (Fig. 2) with symmetrical error-function profiles of electron density across the interfaces, the structure factor can be reduced to the following simple equation:[43-46]

$$\frac{R(q_z)}{R_F(q_z)} \approx F(q_z)\exp(-\sigma^2 q_z^2); \quad F(q_z) = \frac{1}{\rho_b^2}\left|\sum_{m=0}^{2}(\rho_m - \rho_{m+1})\exp(iq_z z_m)\right|^2, \quad (4)$$

where $\sigma = \sigma_0 = \sigma_1 = \sigma_2$, $z_m$ are the locations of the interfaces, $\rho_0 = 0$, the other $\rho_m$ are the electron densities of the layers, and the $\sigma_m$ parameters determine the interfacial width between the slabs of electron density (the standard deviation of their locations, $z_m$).

Overall, fitting the experimental data at $q_z > 0.1$ Å$^{-1}$ either using Parratt formalism or the first Born approximation gives fits with similar quality for layers 1 and 2 ($\rho_1$, $\rho_2$, $\rho_3$, $l_1$, $l_2$ and $\sigma$) that are the same within the error bars of the parameters. The solid lines in Fig. 3 were generated using Eq. (4) with the parameters listed in Table 1; they illustrate the changes in the surface-normal structure after doping the hydrosols with different alkali ions. The only parameter of the surface-normal structure that depends strongly on the alkali metal composition is the density, $\rho_1$. It is noticeably smaller than $\rho_b$ and depends strongly on the dopant's Z-number $Z^+$. In contrast, the thicknesses $l_1$, $l_2$ and the densities $\rho_2$, $\rho_3$ are minimally dependent on the sols' alkali metal composition. I note that the estimated density of the depleted region, $\rho_3$, is close to the electron density of water $\rho_w$. $\sigma$ coincides within the error-bar with the capillary-wave's width $\sigma_{cap} = 2.7 \pm 0.2$ Å, that is given by the detector's resolution, $q_z^{max} \approx 0.7$ Å$^{-1}$ and a short wavelength cutoff in the spectrum of capillary waves: $\sigma_{cap}^2 \approx k_B T \ln(Q_{max}/Q_{min})/(2\pi\gamma)$, where $Q_{min} = q_z^{max} \Delta\beta/2$ and $Q_{max} = 2\pi/a$ ($a \approx 3$ Å is of the order of the intermolecular distance). The surface tension of the sols' surfaces, $\gamma \approx 69-74$ dyn/cm, was measured by a Wilhelmy plate. These results also agree well with the data reported in Ref. [34] for sols with much smaller particles.

Fig. 4a depicts the model distributions of electron density $\rho_1(z)$ in layer 1.[47] Fig. 4b illustrates the dependence of the integral electron-densities of this layer 1, $\Gamma_1 (\sim \rho_1 l_1)$, as a function of $Z^+$, where the circles and squares, respectively, correspond to the x-ray reflectivity data in Fig 3 and in Ref. [34]. The solid line in Fig. 4b is a linear fit of all points. The slope of the line, $\Theta$, is the surface density of alkali ions in layer 1 $\Theta \approx 4 \times 10^{18}$ m$^{-2}$ since $\Theta = d\Gamma_1/dZ^+ \approx (\Gamma_1^M - \Gamma_1^{Na})/(Z_M^+ - Z_{Na}^+)$, where $Z_{Cs}^+$=54, $Z_{Rb}^+$=36, $Z_K^+$=18, and $Z_{Na}^+$=10 are, correspondingly, the numbers of electrons in Cs$^+$, Rb$^+$, K$^+$, and Na$^+$ Accordingly, for the Cs- and Rb-enriched sols the electron density of layer 1 is due to the suspended alkali ions. However, $\Gamma_1 \approx 2 \times 10^{19}$ m$^{-2}$ when $Z^+ \to 0$ (constant term) so that either the alkali ions with small $Z^+$s adsorb in layer 1 with the density, $\Theta$, 50% higher than heavy ions, or the composition of the layer is more complex. For example, there could be one H$_2$O molecule per two alkali ions in the layer (H$_2$0 contains 10 electrons). Indeed, the former suggestion is in the excellent agreement with the grazing incidence diffraction data: at pH=9, the surface density of Na$^+$ is as high as $\Theta_{Na} \approx 6 \times 10^{18} - 7 \times 10^{18}$ m$^{-2}$ [34-35].

The same size effect as in layer 1 is apparent at the surface of the sol enriched by both K$^+$ and Cs$^+$ ions. Since Cs$^+$ is noticeably larger than K$^+$, the former should replace the latter in layer 1 for the same reason that Cs$^+$ (or K$^+$) replaces Na$^+$. The stars and crosses in Fig. 3 correspond to the surface-structure factors of hydrosols containing $\sim$ 0.3 mol/L of Cs and $\sim$ 0.4 mol/L of K, respectively. The estimated integral density of layer 1 of the twice-doped sols is as high as $\Gamma_1 \approx 2 \times 10^{20}$ m$^{-2}$ ($\rho_1 \approx 0.8 \rho_w$ and $l_1 \approx 8$ Å). Then, the content of K$^+$ in layer 1, $x \approx 0.1$, (easily established by solving the following linear equation: $Z_K^+ x + Z_{Cs}^+ (1-x) = \Gamma_1/\Theta$) is in the quantitative agreement with the Ultrup-Kharkats theory.

Thus, we can relate the variation of surface-normal structures of the hydrosols enriched by different alkali metals to the ion-size effect in layer 1: larger ions (for example, $Cs^+$) selectively accumulate in this layer by replacing smaller ions (such as $Na^+$ and $K^+$). However, the estimated center of the layer 1 lies ~ 4 Å above the sol's surface. It is twice as large as the position of the minimum in Fig. 1b. Although the slab model applied in this work was adequate for the spatial resolution of the x-ray reflectivity experiment, $2\pi/q_z^{max}$ ~ 10 Å, it does not afford information at atomic resolution about the true distribution of the ions: increasing the number of layers and/or number of fitting parameters insignificantly improved the quality of the fits. A quantitative interfacial model is required that would take into account, for example, the inhomogeneous spatial distribution of ions along the *z*-axis to make a meaningful comparison between the experimental findings and the Kharkats-Ulstrup theory.

The author would like to acknowledge Anna I. Lygina, Vladimir I. Marchenko, Vitalii V. Zavialov, and Avril Woodhead for valuable discussions and their comments on the manuscript. Beamline X19C received support from the ChemMatCARS National Synchrotron Resource, the University of Chicago, the University of Illinois at Chicago and Stony Brook University. Use of the National Synchrotron Light Source, Brookhaven National Laboratory, was supported by the U.S. Department of Energy, Office of Science, Office of Basic Energy Sciences, under Contract No. DE-AC02-98CH10886. The author also thanks Grace Davison for providing Ludox solutions of colloidal silica.

[47] The distribution of the electron density in the layer 1 can be obtained from the model's electron-density profile

$$\rho(z) \approx \frac{1}{2}\rho_3 + \frac{1}{2}\sum_{m=0}^{2}(\rho_{m+1} - \rho_m)erf\left(\frac{t_m(z)}{\sigma_m\sqrt{2}}\right), \ t_m(z) = z + \sum_{i=0}^{m}z_i, \ \text{and} \ erf(t) = \frac{2}{\sqrt{\pi}}\int_{0}^{t}e^{-s^2}ds,$$

by subtracting from it the contributions of layers 2-3 and the hydrosol's bulk.

**Table 1.** Estimates of the model parameters in Eq. 4 (see also Fig. 2). $c_{Na}^+$ is the bulk concentration of sodium in the hydrosols; $c_M^+$ is the bulk concentration of alkali ions M$^+$ (M = K, Rb, Cs) in the enriched sols; $l_i$ are the thicknesses of the interfacial layers with electron densities $\rho_i/\rho_0$, normalized to the density of bulk water under normal conditions ($\rho_w = 0.333$ e$^-$/Å$^3$); $\sigma_0 = \sigma_1 = \sigma_2 = \sigma$. Parameters $l_1$, $\rho_1/\rho_w$ in the rows shifted upward and downward correspond to the data in Fig. 3 shown by open and solid symbols, respectively. The bulk electron densities of the sols, $\rho_b$, were established from their densities and known chemical compositions. The error bars were estimated utilizing conventional $\chi^2$-criteria at the confidence level 0.95.

| $c_{Na}^+$ (mol/L) | $c_M^+$ (mol/L) | $\rho_b/\rho_0$ | $l_1$ (Å) | $l_2$ (Å) | $\rho_1/\rho_0$ | $\rho_2/\rho_0$ | $\rho_3/\rho_0$ | $\sigma$ (Å) |
|---|---|---|---|---|---|---|---|---|
| 0.1 | - | 1.33 | $8^{\pm 1}$ | $11^{\pm 1}$ | $0.2^{\pm 0.05}$ | $1.20^{+0.08/-0.01}$ | $1.00^{\pm 0.01}$ | $2.8^{\pm 0.2}$ |
| 0.06 | 0.8 (K$^+$) | 1.21 | $7.0^{\pm 0.5}$ / $7.0^{\pm 0.5}$ | $11^{\pm 1}$ | $0.29^{+0.03/-0.04}$ / $0.26^{+0.06/-0.09}$ | $1.26^{+0.02/-0.04}$ | $0.99^{\pm 0.03}$ | $2.7^{\pm 0.3}$ |
| 0.06 | 0.6 (Rb$^+$) | 1.24 | $7.7^{\pm 0.5}$ / $7.6^{\pm 0.5}$ | $11.8^{\pm 0.5}$ | $0.84^{+0.05/-0.04}$ / $0.51^{+0.05/-0.07}$ | $1.30^{+0.04/-0.03}$ | $1.07^{\pm 0.03}$ | $2.7^{\pm 0.2}$ |
| 0.06 | 0.7 (Cs$^+$) | 1.24 | $9.0^{\pm 0.3}$ / $6.5^{\pm 0.4}$ | $11.4^{\pm 0.5}$ | $0.93^{\pm 0.04}$ / $0.90^{\pm 0.04}$ | $1.31^{\pm 0.04}$ | $1.05^{\pm 0.05}$ | $2.8^{\pm 0.3}$ |

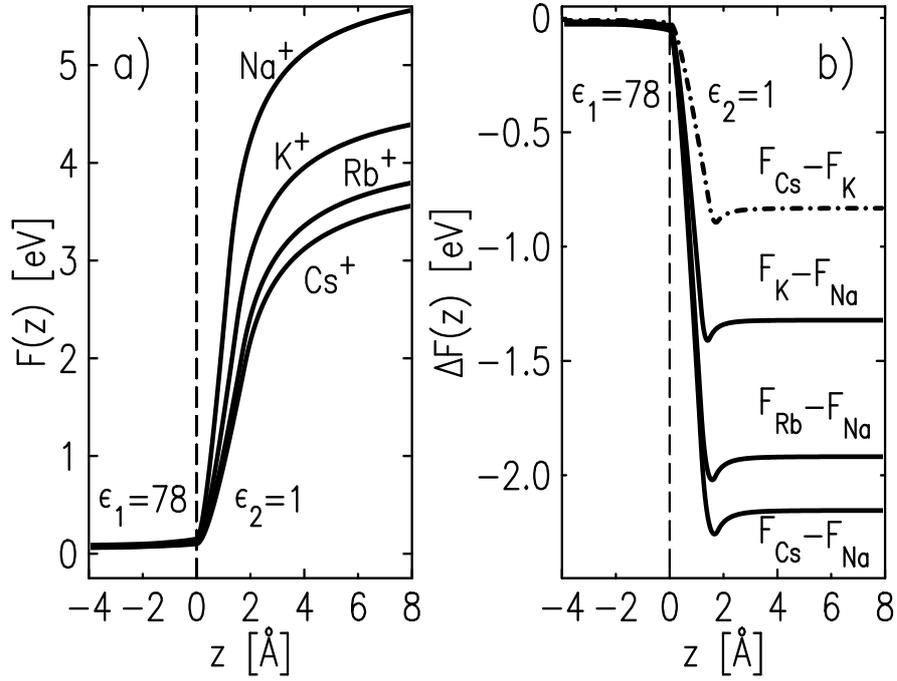

**Figure 1**. Kharkats-Ulstrup size-effect at the air-water interface: a) single-ion electrostatic free energy of the alkali ions at the air-water interface as a function of $z$; b) differences between free energies, $F_M(z)$, of monovalent alkali ions (M=Na, K, Rb, Cs) at the air-water interface. For Na$^+$ the radius $a \approx 1.2$ Å; for K$^+$ $a \approx 1.5$ Å; for Rb$^+$ $a \approx 1.7$ Å; and, for Cs$^+$ $a \approx 1.8$ Å.[30-31]

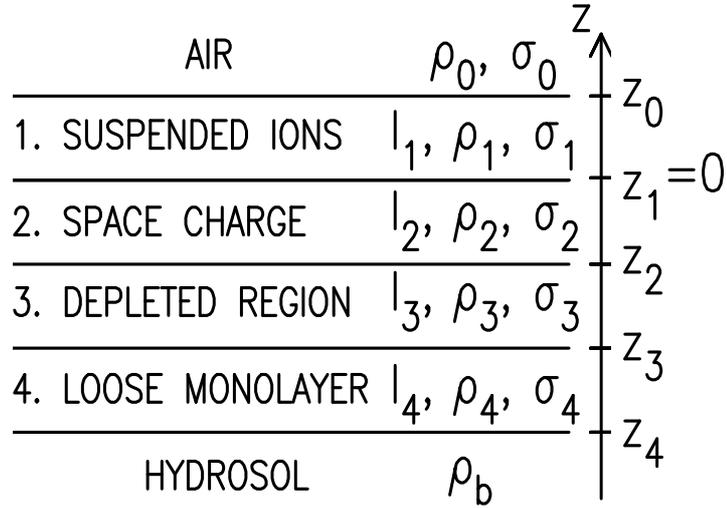

**Figure 2**. The four-layer slab model of hydrosol's surface-normal structure. Each layer has a thickness $l_m$, and a homogeneous electron density $\rho_m$. In addition, $\sigma_m$ parameters determine the interfacial width between slabs of electron density (the standard deviation of their locations $z_m$). At $q_z > 0.1$ Å$^{-1}$ only three interfaces (top two layers with adsorbed ions) contribute to the reflectivity since $\sigma_3 \sim \sigma_4 \gg \sigma_0, \sigma_1, \sigma_2$. The only parameter of the surface-normal structure that depends strongly on the alkali metal composition is the density, $\rho_1$.

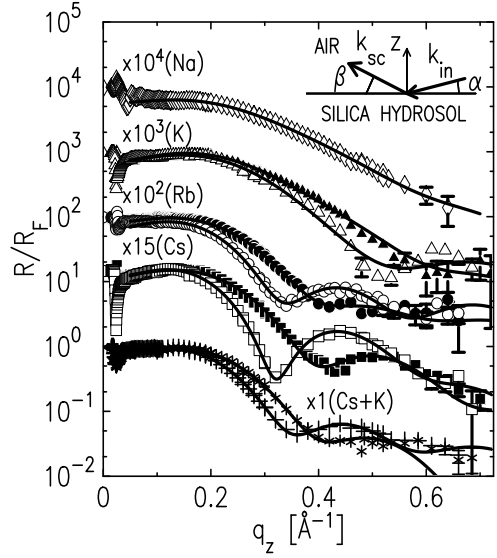

**Figure 3**. The surface structure factors of the 22-nm-particle sols: the rhombi represent sol stabilized by NaOH, $c_{Na}^+ \approx 0.1$ mol/L; the filled and open triangles are for potassium-enriched sols with $c_K^+ \approx 0.8$ mol/L and $c_{Na}^+ \approx 0.06$ mol/L; the dots and circles are for rubidium-enriched sols with $c_{Rb}^+ \approx 0.6$ mol/L and $c_{Na}^+ \approx 0.06$ mol/L; the filled and open squares are for cesium-enriched sols with $c_{Cs}^+ \approx 0.7$ mol/L and $c_{Na}^+ \approx 0.06$ mol/L. Here, filled and open symbols on each R/R$_F$ curves refer to samples with different equilibration history. The crosses and stars are for mixtures of cesium- and potassium- enriched sols with $c_K^+ \approx 0.4$ mol/L, $c_{Cs}^+ \approx 0.3$ mol/L and $c_{Na}^+ \approx 0.06$ mol/L. The lines denoted the first Born approximation that is discussed in the text. Insert: $\mathbf{k}_{in}$ and $\mathbf{k}_{sc}$ are, respectively, wave-vectors of the incident beam, and beam scattered towards the point of observation, and $\mathbf{q}$ is the wave-vector transfer, $\mathbf{q} = \mathbf{k}_{in} - \mathbf{k}_{sc}$. At reflectivity conditions ($\alpha = \beta$) there is only one component of the wave-vector transfer, $q_z = (4\pi/\lambda)\sin(\alpha)$, where $\alpha, \beta$ are the angles of the incident- and scattered-beams in the plane normal to the surface. The reflectivity was measured with the detector's vertical slits gap ~ 0.8 mm at the distance ~700 mm form the footprint or angular acceptance at $\Delta\beta = 6.8 \times 10^{-2}$ deg (twice higher than in Ref. [34]) and its horizontal acceptance at ~ 0.8 deg (~ 10-mm-gap).

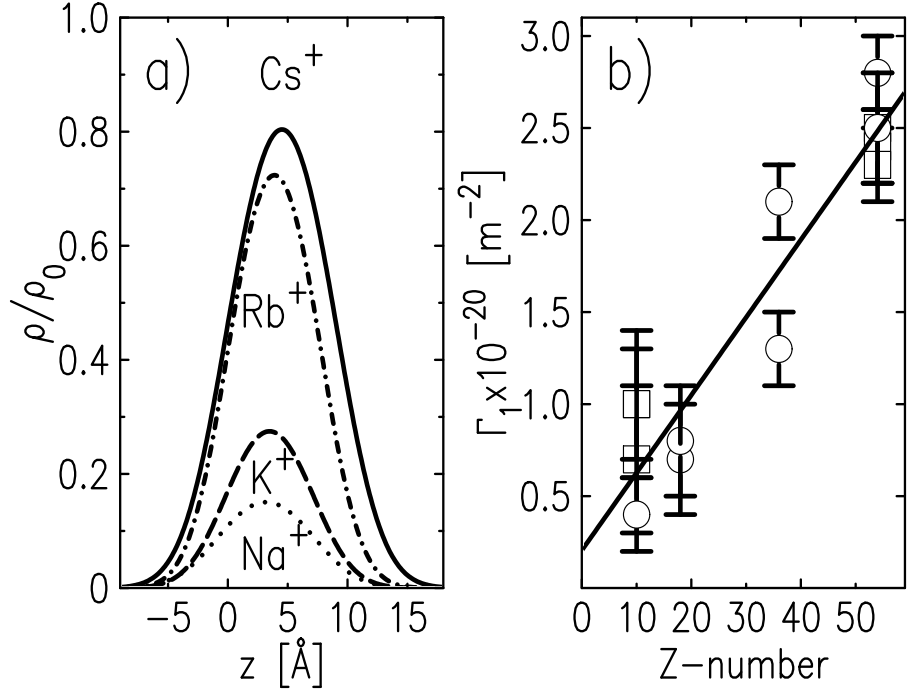

**Figure 4**. a) Model distributions of electron density $\rho_1(z)$ in layer 1. b) Integral density of layer 1 vs. the number of electrons in the alkali ion, where $Z^+_{Cs}=54$, $Z^+_{Rb}=36$, $Z^+_{K}=18$ and $Z^+_{Na}=10$ are, respectively, the numbers of electrons in $Cs^+$, $Rb^+$, $K^+$, and $Na^+$. The circles and squares correspond to the reflectivity curves in Fig 3 and the data obtained from Ref. 34, respectively. A solid line is the linear fit of these data.